\documentclass[doublecol]{epl2}
\usepackage{graphicx}
\usepackage{dcolumn}
\usepackage{bm}% bold math

\usepackage{colordvi}

\begin{document}

\title{Manganese-diffusion-induced n-doping in semiconductor structures containing  Ga(Mn)As layers}
\shorttitle{Mn-diffusion-induced n-doping in structures containing
Ga(Mn)As layers}
\author{T. Korn\inst{1} \and R. Schulz\inst{1}
\and S. Fehringer\inst{1} \and U. Wurstbauer\inst{1} \and D.
Schuh\inst{1} \and W.\ Wegscheider\inst{1} \and M. W. Wu\inst{2}
\and C.\ Sch\"uller\inst{1}} \institute{
  \inst{1} Institut f\"ur Experimentelle und Angewandte Physik,
Universit\"at Regensburg, D-93040 Regensburg, Germany\\
  \inst{2} Hefei National Laboratory for Physical Sciences at
Microscale and Department of Physics, University of Science and
Technology of China, Hefei, Anhui, 230026, China }
\shortauthor{T.
Korn \etal}
 {Electrical properties of specific thin
films - III \- V semiconductors }
 {Time resolved
luminescence }

\abstract{Semiconductor structures using ferromagnetic
semiconductors as spin injectors are promising systems for future
spintronic devices. Here, we present combined photoluminescence (PL)
and time-resolved magneto-optical experiments of nominally
nonmagnetic quantum wells (QWs) separated by a thin barrier from a
ferromagnetic Ga(Mn)As layer. Due to the partial quenching of the
PL, we conclude that there is a significant Mn backdiffusion into
the QW. Moreover, from the time-resolved measurements, we infer that
the Mn leads to n-type doping within the QW, and, in addition,
strongly increases the electron spin dephasing time. The amount of
Mn backdiffusion strongly depends on the barrier composition.}

\maketitle One key requirement for the development of semiconductor
spintronic devices is the reliable injection of spin-polarized
carriers into a nonmagnetic semiconductor structure
\cite{Fabian04,Fabian07}. Here, ferromagnetic semiconductors like
Ga(Mn)As \cite{Ohno1996} have emerged as an alternative to metallic
ferromagnets.  Highly efficient spin injection from Ga(Mn)As into a
nonmagnetic heterostructure was demonstrated using a spin-LED
structure \cite{VanDorpe04}. Additionally, heterostructures using
Ga(Mn)As show new magnetotransport phenomena, both tunneling
anisotropic magnetoresistance (TAMR) \cite{Ruster05}, where a single
ferromagnetic layer produces a spin-valve-like signal, and
large-amplitude tunneling magnetoresistance (TMR) \cite{Gould04}. In
the TMR structures, a thin (2~nm) GaAs layer between two Ga(Mn)As
layers serves as a tunnel barrier.

In order to grow ferromagnetic Ga(Mn)As layers by molecular beam
epitaxy (MBE), very low growth temperatures have to be used to
inhibit Mn diffusion and MnAs formation. Due to the strong diffusion
and segregation of Mn, it is expected that in any semiconductor
structure containing a Ga(Mn)As layer, Mn will also be incorporated
above and below the Ga(Mn)As. This is likely to cause background
doping and strongly influences the transport properties of adjacent
layers. Due to the close proximity of the p-metallic Ga(Mn)As,
however, it is impossible to independently determine the type and
concentration of this background doping by transport measurements.

Here, we investigate this background doping by time- and
spectrally-resolved optical techniques using a model system, which
represents the combination of a ferromagnetic semiconductor and a
two-dimensional structure. The combined analysis of time-resolved
photoluminescence (TRPL) and time-resolved Faraday/Kerr rotation
(TRFR/TRKR) experiments allows us to draw the conclusion that in our
structures Mn diffusion leads to an n-type doping.

Our sample structure consists of two nominally undoped
GaAs/Al$_{0.3}$Ga$_{0.7}$As quantum wells (QW) of different width,
grown on a [001] GaAs substrate at the typical growth temperature of
$600^o$C. The narrower QW (10~nm, here referred to as ref. QW) is
grown first and serves as a reference. It is well-separated from the
structures above by a 100~nm Al$_{0.3}$Ga$_{0.7}$As barrier. The
wider QW (12~nm, here called Mn QW) is separated by a thin barrier
from the 50~nm thick Ga(Mn)As layer, which is grown last at low
temperature ($250^o$C). Both, the QWs and the barrier  are grown in
the high-mobility chamber of a two-chamber MBE system (capable of
achieving two-dimensional electron system mobilities well above
$10^7$cm$^2$/Vs). The sample is then transferred in vacuum to the
second chamber, where the Ga(Mn)As layer is grown. Note that this
sample design is different from systems studied by Poggio et al.,
where Mn was directly incorporated into a QW during growth, causing
incorporation of Mn on Ga sites \cite{Poggio05}. Here, we present
results from three samples with different barrier thickness and
composition. Samples A and B: 4.34~nm of Al$_{0.3}$Ga$_{0.7}$As is
grown on top of the QW layer. Then five periods of 3 monolayers (ML)
AlAs / 1 ML GaAs are grown without rotation of the wafer, resulting
in a total barrier thickness of 8.8~nm (sample A) and 11.1~nm
(sample B). All layers are grown at $600^o$C. Sample C: 5~nm of
Al$_{0.3}$Ga$_{0.7}$As is grown on top of the QW at $600^o$C,
followed by 5~nm of Al$_{0.8}$Ga$_{0.2}$As grown at $250^o$C. We
note that in all three samples, the total barrier thickness is
significantly larger than in  TMR structures based on Ga(Mn)As/GaAs.
\begin{figure}
  \includegraphics[width= 0.4\textwidth]{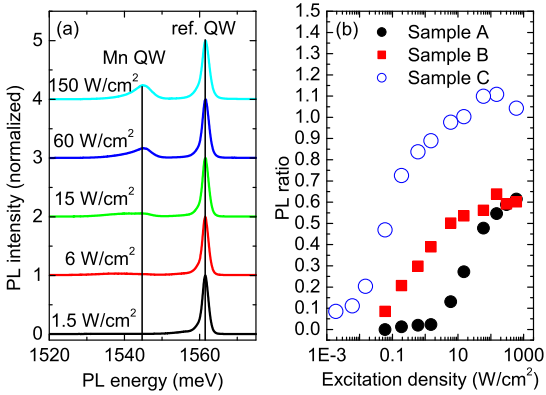}
   \caption{(a) PL traces for the two QWs for different excitation densities,
   measured on sample A.
   (b) ratio of the PL peak areas for the two QWs as a function of excitation
density. Black dots indicate sample A, red squares sample B, and
blue open circles sample C.}
   \label{PL_2Panel}
\end{figure}

We first study the photoluminescence (PL) of these samples as a
function of excitation density, using a green (532~nm) cw laser.
Figure \ref{PL_2Panel} (a) shows typical PL spectra of sample A
measured at 4~K. For low excitation density, only the PL of the ref.
QW is visible. As the excitation density is increased, a PL signal
from the Mn QW appears, but comparison between the  PL lineshapes
shows a significant broadening, and a pronounced low-energy shoulder
in the emission from the Mn QW. In a reference sample grown in the
same manner, but with the top layer consisting of
low-temperature-grown GaAs without any Mn content, both QWs show
essentially the same PL linewidth. As in a previous study
\cite{SchulzPE}, where we investigated different barrier widths,
grown at high temperatures, we attribute this broadening and the
quenching at low excitation density to Mn backdiffusion into the QW
during growth of the Ga(Mn)As layer. In the QW, Mn ions act as
centers for nonradiative recombination and partially quench the PL.
Additionally, they also allow for the formation of bound excitons,
which are redshifted by the additional binding energy. The
observation of the broad low-energy shoulder in the PL spectrum,
instead of a  peak at low energy corresponding to a well-defined
bound exciton state, indicates a large inhomogeneity of the bound
states. Sample B and C show the same behavior, however, the PL from
the Mn QW appears at lower excitation density. In Figure
\ref{PL_2Panel} (b), we show the ratio of the PL intensities of the
Mn QW and the ref. QW, calculated from the peak areas, for all three
samples. It is clearly visible how the PL from the Mn QW appears at
a threshold excitation power and then saturates as the excitation
power is increased further. In samples A and B, it only reaches
about 60~percent of the ref. QW intensity, while in sample C both
QWs have comparable PL intensity at high excitation. This behavior
indicates that the nonradiative recombination centers are saturated
as the density of photocarriers is increased. The different
threshold powers of the three samples show that Mn diffusion depends
strongly on the barrier thickness, and that a high Al content/low
temperature-grown barrier is significantly more efficient at
suppressing Mn diffusion than a superlattice. In order to observe
the influence of Mn on the spin dynamics in the QW, we mainly focus
on sample A, where the effects are more pronounced due to the
largest Mn content.
\begin{figure}
  \includegraphics[width= 0.4\textwidth]{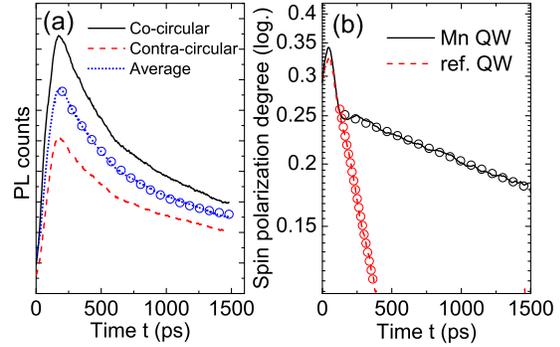}
  \caption{(a) TRPL traces of the Mn QW in sample A for co- and
  contracircular configuration.
  The dotted line is the average of both. (b) Spin polarization degree traces for
  Mn  and ref. QW in sample A. Open circles in both panels indicate exponential fits
  to the data.}
   \label{TRPL_2Panel}
\end{figure}

We next discuss TRPL measurements performed at 4~K. Here, the sample
is nonresonantly excited by circularly polarized 600~fs pulses from
a modelocked Ti:sapphire laser. The circular polarization degree of
the resulting PL is analyzed by a quarter-wave-plate and a
polarizer. The PL is then detected by a streak camera system. Both,
the recombination dynamics of the photocarriers in either QW, and
their spin dynamics can be extracted from the TRPL traces. In
subsequent measurements, first the TRPL with the same helicity
(co-circular) as the excitation laser is recorded, then the TRPL
with the opposite helicity (contra-circular). Figure
\ref{TRPL_2Panel}(a) shows two time traces generated from TRPL data
by averaging over a spectral window of 10~meV width around the
spectral peak of the Mn QW. It is clearly visible that the
co-circular TRPL is stronger than the contra-circular trace in the
whole time window, indicating a finite spin polarization of the
photocarriers. The dotted line between the traces corresponds to the
average of the traces. This reflects the recombination dynamics of
the photocarriers. After a pronounced maximum of the TRPL, it decays
exponentially with a time constant of 490~ps. For the ref. QW, a
similar decay constant of the TRPL is found. The spin polarization
degree (SPD) of the TRPL is calculated as follows: the difference
between the co- and contra-circular traces is divided by their sum.
Figure \ref{TRPL_2Panel}(b) shows the SPD of the Mn and the ref. QWs
as a function of time on a logarithmic scale. While in the ref. QW,
the SPD decays more quickly than the TRPL (313~ps decay time), we
observe a significantly longer SPD decay in the Mn QW (3900~ps). One
explanation for this drastic increase in spin dephasing time is
motional narrowing: in one regime of the D'Yakonov-Perel spin
dephasing mechanism \cite{Dyakonov72}, which is dominant in GaAs
heterostructures at low temperatures, the spin dephasing time is
inversely proportional to the momentum relaxation time. In the Mn
QW, the Mn ions are efficient momentum scattering sites, leading to
shorter momentum relaxation time than in the ref. QW. Although
electron scattering at paramagnetic impurities like Mn allows for
electron spin flips, at low impurity density  motional narrowing
dominates \cite{Wu}. Additionally, electron localization at defects
may also contribute to the increased spin dephasing time.
\begin{figure}
 \includegraphics[width= 0.4\textwidth]{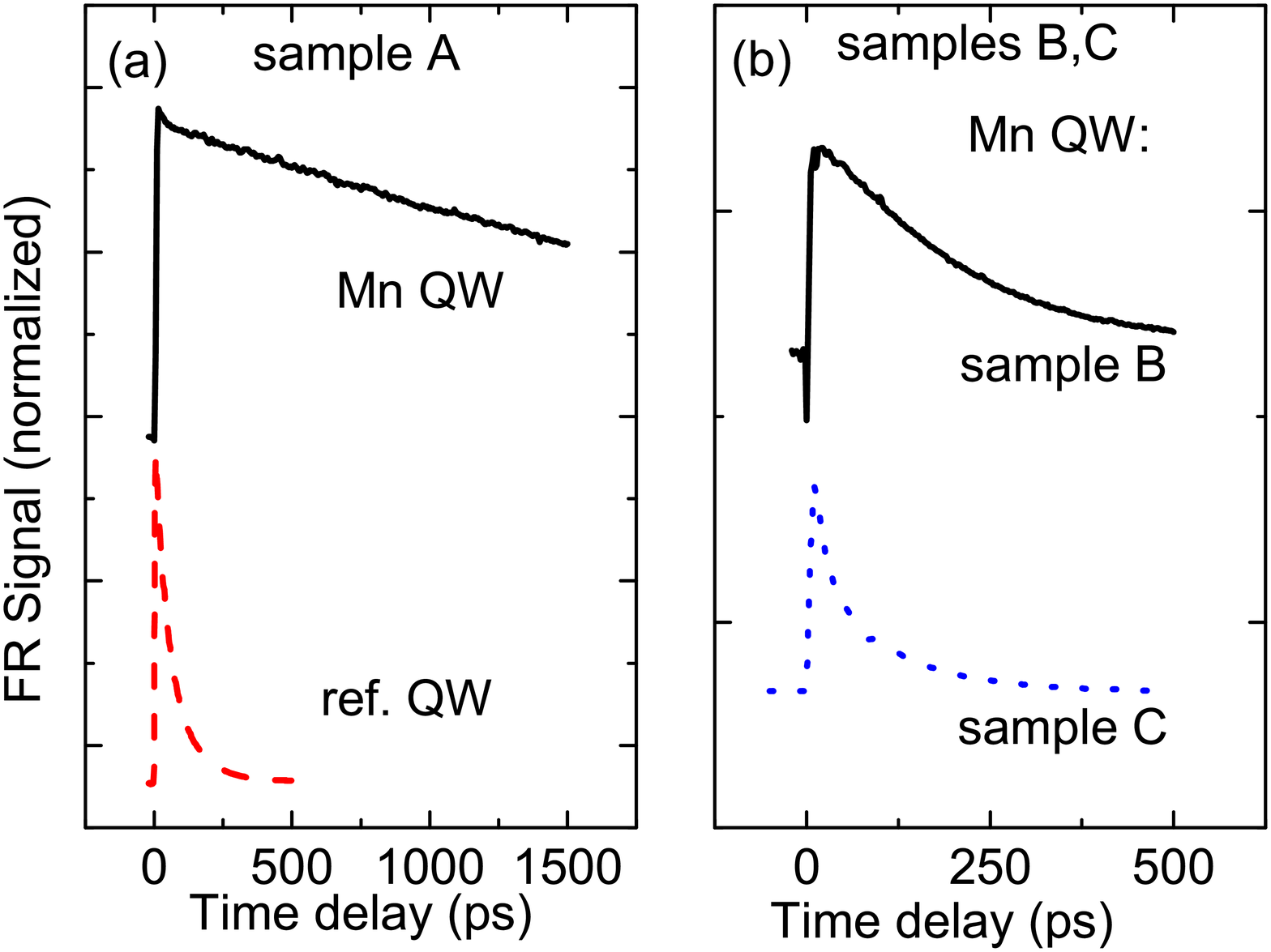}
  \caption{(a) TRFR traces of the Mn QW (black solid line) and the ref. QW (red dashed line)
   in sample A, taken at zero magnetic field.
  (b) TRFR traces of the Mn QWs in samples B (black solid line) and C (blue dotted line),
  taken at zero magnetic field.
  Note the different time scales in (a) and (b).}
\label{TRFR_VGL}
\end{figure}

We further investigate this increase in spin dephasing time by
performing TRFR/TRKR measurements. Both are two-beam pump-probe
techniques: a circularly-polarized pump pulse from a mode-locked
Ti:Sapphire laser is used for near-resonant excitation of either the
Mn or the ref. QW. The spin polarization in the QWs is then detected
by a second, time-delayed probe pulse from the same laser. This
probe pulse is linearly polarized. Its polarization plane is rotated
by a small angle proportional to the spin polarization, due to the
Faraday (transmission) or Kerr (reflection) effects. This  rotation
angle is analyzed by an optical bridge detector. By scanning the
delay between pump and probe pulses, the time dependence of the spin
polarization is tracked. For measurements in transmission, the
samples are glued to a transparent sapphire substrate, and the wafer
side of the samples is removed by mechanical grinding and wet
chemical etching, leaving only the MBE-grown layers.

Some differences between TRPL and TRFR/TRKR measurements need to be
noted: in TRPL, nonresonant excitation is used, and spin information
can only be extracted from the sample during the duration of the PL.
Due to the selection rules in a GaAs QW and the fact that holes
typically lose their spin orientation within a few ps
\cite{Damen91}, the circular polarization of the PL arises from the
recombination of spin-polarized electrons. However, the spin
polarization degree of the PL does not reflect the absolute spin
polarization within the sample, which may change considerably due to
carrier recombination: If we consider an undoped sample in which
there is no electron spin relaxation, the TRPL spin polarization
degree may remain constant during the duration of the PL, while the
spin polarization in the sample decreases to zero as the
photocarriers recombine. In TRFR/TRKR, resonant excitation is used
to selectively excite one of the QWs, and the Faraday/Kerr rotation
(FR/KR) signal is proportional to the absolute spin polarization
within the QW. This spin polarization may arise from photocarriers,
resident electrons or holes, or even localized moments due to
magnetic impurities. Therefore, if there is a coherent spin
polarization of resident carriers in the QW due to background
doping, even after photocarrier recombination, a finite signal is
measured.
\begin{figure}
 \includegraphics[width= 0.4\textwidth]{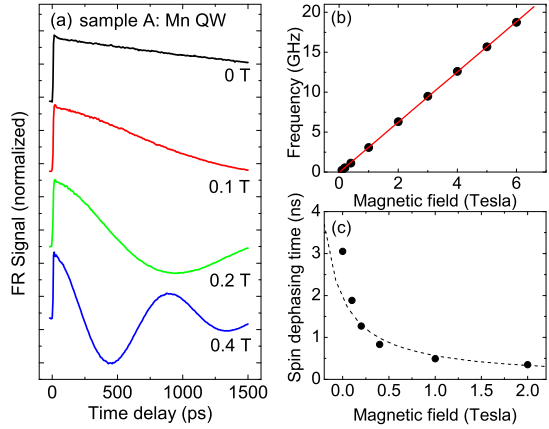}
  \caption{(a) TRFR traces of the Mn QW in sample A for different
  applied in-plane magnetic fields.
  (b) Spin precession
  frequency for different in-plane magnetic fields measured for the Mn QW
  in sample A. The red line represents a linear fit.
  (c) Spin dephasing time of the Mn QW in sample A as a function of magnetic field.
  The dotted line represents a 1/B dependence.}
\label{TRFR_mag}
\end{figure}

Figure \ref{TRFR_VGL}(a) shows  TRFR traces taken at 4~K, where the
laser was tuned to be resonant with the ground-state transitions of
either the Mn or the ref. QW in  sample A. Under these excitation
conditions, the photocarrier recombination time is typically much
faster than for nonresonant excitation \cite{Zhukov}. We observe a
fast decay of the spin polarization for the ref. QW in sample A of
about 100~ps, reflecting the photocarrier lifetime in the nominally
undoped structure. Similar results are seen for the ref. QWs in all
three samples.  For the Mn QW in sample A, we observe a
significantly longer decay of  about 3~ns, corresponding to the slow
decay of the SPD observed in the TRPL measurements, and indicating
that a resident spin polarization remains within the QW after
photocarrier recombination. In sample B (see Figure
\ref{TRFR_VGL}(b)), the spin lifetime of the Mn QW (200~ps) is also
longer than that of the ref. QW (100~ps), while for sample C we
observe similar spin lifetimes for both, the Mn QW and the ref. QW.
These observations correspond to the intensity-dependent PL data
(Fig. \ref{PL_2Panel} (b)), in which sample C showed the lowest
threshold excitation density and therefore the lowest Mn content.

In order to identify the type of the resident carriers, we apply an
in-plane magnetic field to induce spin precession. In Figure
\ref{TRFR_mag}(a), we show TRFR traces of the Mn QW (sample A) for
different in-plane magnetic fields. A damped oscillation is
observed, and both, the frequency and the damping of the
oscillation, increase with magnetic field. A damped cosine function
is fitted to the data  to extract the frequency and the damping
constant. In figure
 \ref{TRFR_mag}(b), this frequency is plotted as a function of the magnetic field.
A linear fit to this dependence yields the g factor, $|g|=0.225$.
This value is in good agreement with the electron g factor in QWs of
similar width \cite{Snelling91}. From this, we can identify the
resident carriers in the QW as electrons. The presence of resident
electrons in the QW indicates that the Mn ions, which diffuse from
the Ga(Mn)As layer into the heterostructure  during growth, are
incorporated mostly as interstitials, where they act as double
donors \cite{Maca02}. In figure \ref{TRFR_mag}(c) we plot the spin
dephasing time as a function of the in-plane magnetic field. A
strong 1/B-like dependence is seen, indicated by the dotted line.
This behavior is caused by an inhomogeneity $\Delta g$ of the
electron g factor within the sample.  The likely cause for this
strong inhomogeneity is electron localization at defects
\cite{Chen07}.

In conclusion, we have investigated the diffusion of Mn ions from
ferromagnetic Ga(Mn)As layers into a nonmagnetic heterostructure by
optical spectroscopy techniques.  The combination of time-resolved
photoluminescence and time-resolved Faraday rotation allows us to
determine that the diffusing Mn ions are incorporated as
interstitials, and therefore act as double donors, causing an
n-doping of layers adjacent to Ga(Mn)As. Additionally, the presence
of Mn ions strongly increases the electron spin lifetime in the
heterostructure. The diffusion of Mn strongly depends on the barrier
composition and thickness. It can be partially suppressed by using a
low-temperature-grown barrier with high Al content. Support by the
DFG via SPP 1285 and SFB 689, by the Natural Science Foundation of
China under Grant 10725417, and the Robert-Bosch-Stiftung is
gratefully acknowledged.

\end{document}